\documentclass[twocolumn,showpacs,amsmath,amssymb]{revtex4}
\input epsf
\begin{document}
\title {2-Dimensional Dipolar Scattering}
\author{Christopher Ticknor}
\affiliation{ARC Centre of Excellence for Quantum-Atom Optics and
Centre for Atom Optics and Ultrafast Spectroscopy,
Swinburne University of Technology, Hawthorn, Victoria 3122, Australia }
\affiliation{ITAMP, Harvard-Smithsonian Center for Astrophysics,
Cambridge, Massachusetts 02138, USA}
\date{\today}
\begin{abstract}
We characterize the long range dipolar scattering in 2-dimensions.  
We use the analytic zero energy wavefunction including the dipolar 
interaction; this solution yields universal dipolar scattering properties in 
the threshold regime.  We also study the semi-classical dipolar scattering and
find universal dipolar scattering for this energy regime. 
For both energy regimes, we discuss the validity of the universality and give
physical examples of the scattering.
\end{abstract} 
\maketitle

Many theoretical proposals are based on dipolar gases in
2-dimensional (2D) geometries. For example, there are predictions of 
Wigner crystals \cite{buchler}, anisotropic solitons \cite{tik}, and phonon
instabilities \cite{nath}.  Such theories show dipolar systems will lead to 
exotic and highly correlated quantum systems.  Reduced  dimensionality 
also offers another level of control to exert over ultracold matter.  
In pursuit of such quantum 
systems, there has been exciting experimental progress in the production of 
polar molecules \cite{gspm,carr} and in the production of quasi-2 
dimensional ultracold gases \cite{bkt,2D}.  This makes it seem that the 
experimental production of 2D dipolar gases is at hand.
However, there is no simple understanding of dipolar scattering in 2D, even for
the case of the long range scattering.  Such an understanding is an important
first step in the study of these quantum systems. 
In this paper we study the properties of long range dipolar scattering in 2D 
and present estimates of the scattering cross section.
We consider the scenario when the polarization of the molecules 
($\hat d=\hat z$) is perpendicular to the plane of motion 
($\rho=\sqrt{x^2+y^2}$,  $\hat d\cdot\hat\rho=0$).  
We do not consider the complications of transverse 
confinement.  Under these simplifications the dipolar interaction is: 
$V_{dd}=d^2/\rho^3$, where $d$ is the magnitude of the induced dipole moment.  
The anisotropy of the interaction has been removed and the interaction is 
purely repulsive.  

To understand the scattering we use the dipolar length scale: 
$D=\mu d^2/\hbar^2$ where $\mu$ is the reduced mass.
In polar molecules this length scale can be quite large, orders of magnitude
larger than the range of the short range interaction, $\rho_0$.  
Using $D$ to rescale the 2D radial Schr\"{o}dinger equation 
and expanding in partial waves, 
$\psi(\tilde\rho,\varphi)=\sum_me^{im\varphi}\phi_m(\tilde\rho)/\sqrt{\tilde\rho}$, the result is:
\begin{eqnarray}
\left(-{d^2\over d\tilde\rho^2}+{m^2-1/4\over \tilde\rho^2}+
{2\over \tilde\rho^3}\right)\phi_{m}(\tilde\rho)={(Dk)^2}\phi_m(\tilde\rho),
\label{TISE}
\end{eqnarray}
where $\tilde\rho=\rho/D$, $k^2=2\mu E/\hbar^2$, and $E$ is the scattering 
energy.  If $\tilde\rho=\rho_0/D\ll1$ then the only degree of freedom is $Dk$, 
so once Eq. (\ref{TISE}) is solved for a given $Dk$ the resulting scattering 
will apply to any quantum mechanical 2D dipolar system; 
this is universal dipolar scattering \cite{universal,roudnev,NJP}.

We have analyzed Eq. (\ref{TISE}) and characterized the long range 2D dipolar 
scattering. We present analytic estimates of the scattering in both the 
threshold and semi-classical limits.
In contrast to 3D, there is a diagonal s-wave 
dipolar interaction, and this leads to universal dipolar scattering in the 
threshold regime.  We refer to the isotropic $m=0$ partial wave as s-wave.
For threshold scattering in 3D, it is required that $a_s\ll D$ to have 
universal dipolar scattering \cite{roudnev,NJP}.  This is challenging because 
$a_s$, the s-wave scattering length, depends sensitively on the details of 
the short range interaction.  However in 2D, the repulsive dipolar 
interaction prevents ultracold particles from reaching the short range 
interaction where it can acquire a non-universal phase shift.  To have 
universal dipolar scattering we require only that the range of the short 
range interaction is much smaller than the dipolar length scale 
($\tilde\rho_0\ll 1$). This criteria does not rely on serendipity to be met.

In this regime, the s-wave scattering properties are derived from an analytic 
zero energy solution to the Schr\"{o}dinger equation {\it including} the 
dipolar interaction.  This solution shows that dipolar interaction results in
an s-wave scattering length of 6.344$D$.  We also present the Born approximation
which predicts the threshold scattering for non-zero partial waves.
In the large $Dk$ limit, the {\it semi-classical universal} dipolar 
scattering cross section can be estimated by the Eikonal approximation and 
scales as $\sqrt{D/k}$.  These estimates offer accurate predictions of the
scattering in their respective energy regimes.

To monitor the scattering we use the scattering cross section, which in 2D 
for distinguishable particles is
$\sigma={4\over k}\sum_{m=-\infty}^{\infty}\sin^2(\delta_{m})$, 
where $\delta_m$ is the 
scattering phase shift for the $m^{th}$ partial wave \cite{cross,adhikari,gu}. 
For identical bosons (fermions) the sum is restricted to even (odd) $m$
and there is an additional factor of 2. 
Note that in 2D, $\sigma$ has units of length.
We numerically solve Eq. (\ref{TISE}) and
obtain the scattering phase shift by matching $\phi_m$ to the free  
solution at large $\rho$: 
$\sqrt{k\rho}[\cos(\delta_{m})J_m(k\rho)-\sin(\delta_{m})N_m(k\rho)]$,
where $J_m/N_m$ are regular Bessel functions of order $m$.
An interesting fact of 2D scattering is that the s-wave, 
cross section diverges as $k\rightarrow0$. 
To illustrate this, we solve the s-wave scattering 
with no dipolar interaction and a hard wall at $\rho_0$. This yields
\begin{eqnarray}
&&\sigma_{hw}(\rho_{0})={4\over k}\sin^2(\delta_{hw}),\\
&&\tan(\delta_{hw})=J_{0}(k\rho_0)/N_0(k\rho_0).\nonumber
\end{eqnarray}
As $k$ goes to zero the phase shift becomes logarithmically small because 
$N_0\propto\ln(k\rho_0)$; however this is not quick enough to counter 
the diverging $1/k$. 

\begin{figure}
\vspace{2mm} 
\centerline{\epsfysize=50mm\epsffile{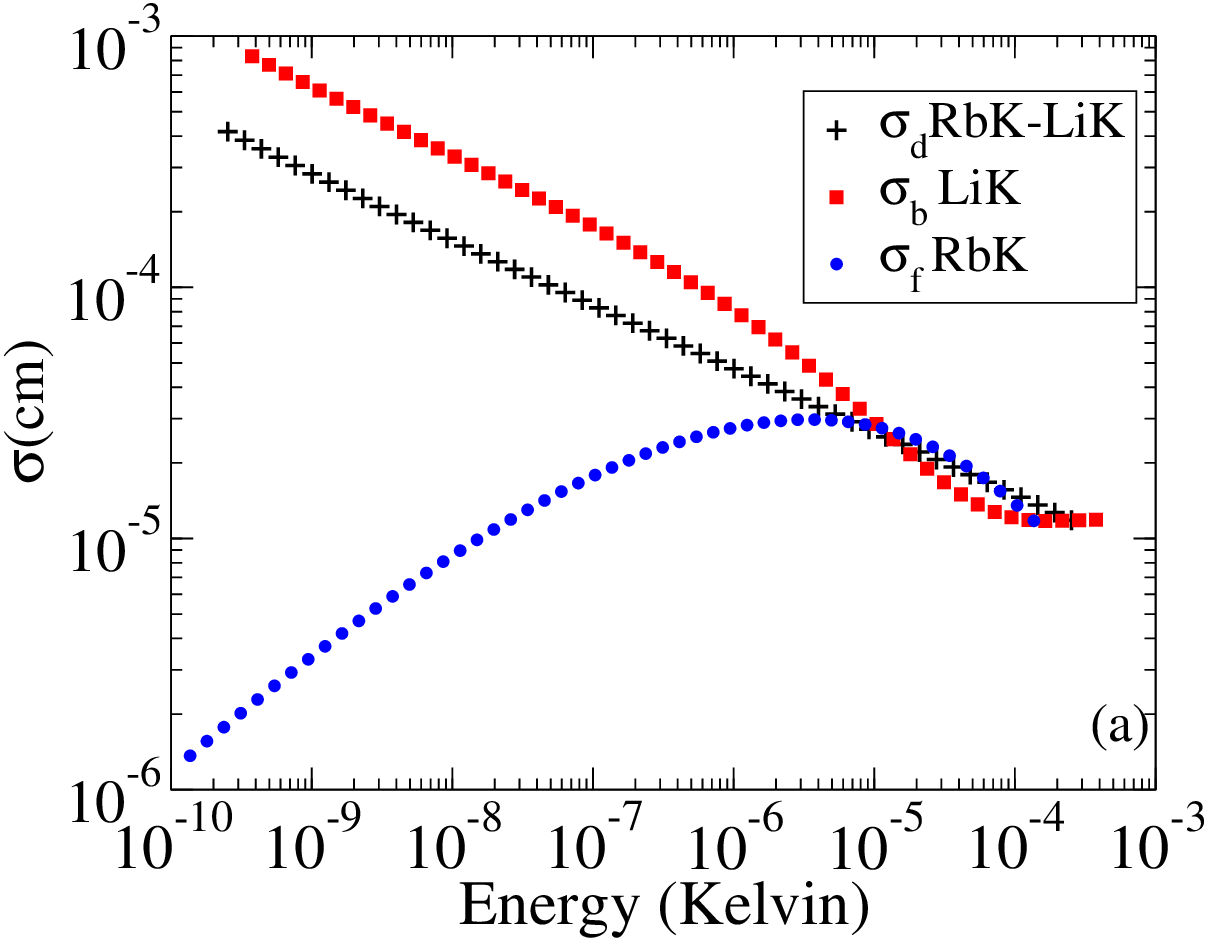}} 
\vspace{2mm} 
\centerline{\epsfysize=50mm\epsffile{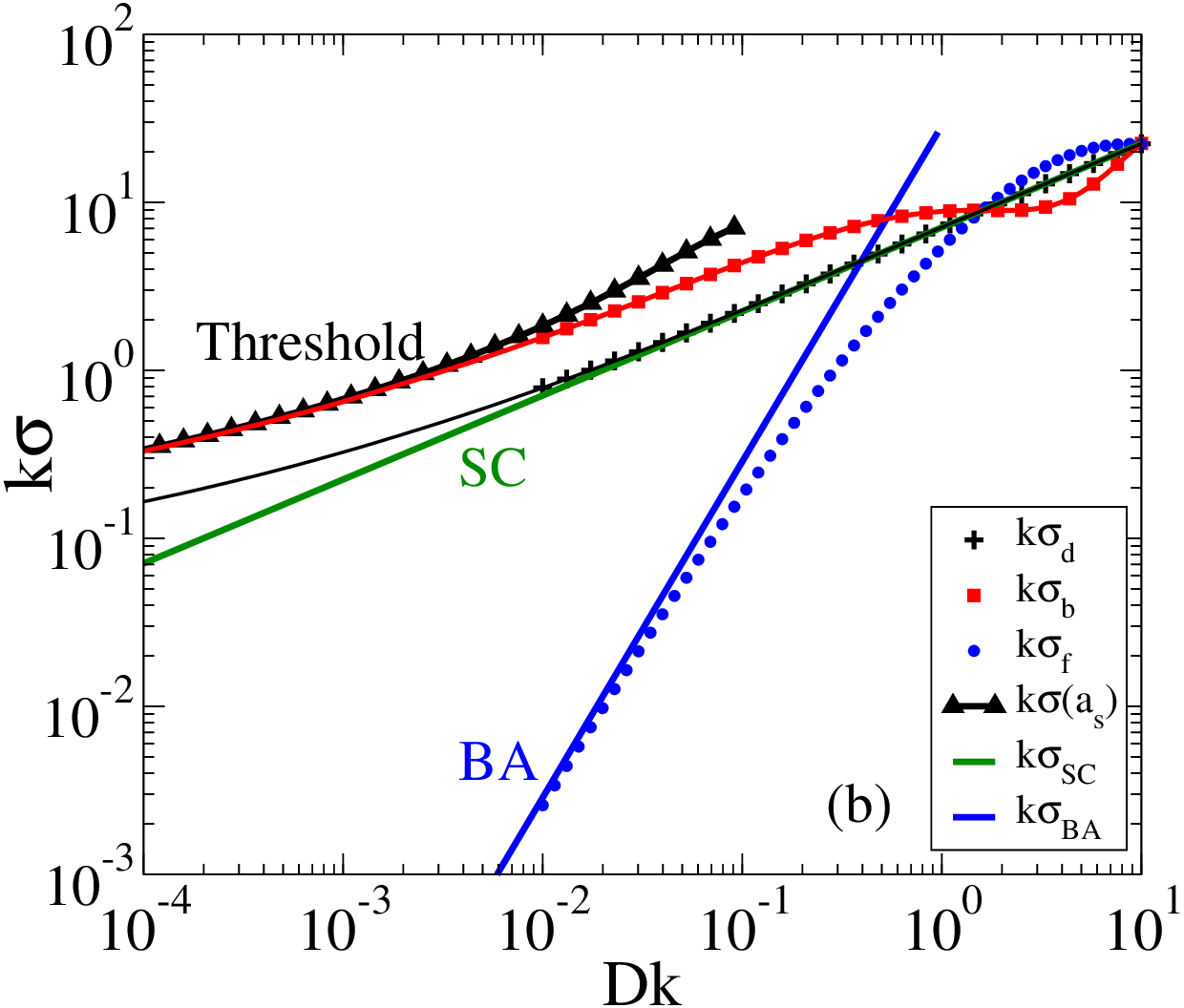}} 
\caption{(Color Online) (a) The 2D scattering cross section is shown for 
bosonic KLi (red squares, $\sigma_b$),  
fermionic RbK (blue circles, $\sigma_f$),
and distinguishable RbK-KLi (black $+$, $\sigma_d$) for 
$D=5.29\times10^{-6}$cm.  
(b) The dimensionless $k\sigma$ is shown with  
three analytic estimates; they are the
Born Approximation (blue solid line), semi-classical (solid green line), 
and the analytic s-wave threshold scattering.
The symbols are data from (a) with their energy range extended to 
show the low energy trend of the scattering.}\label{sigma}
\end{figure}

We have plotted the cross section for three physical systems.  
In Fig. \ref{sigma}(a) we show the cross section as a function of energy
for bosonic LiK ($\sigma_b$), fermionic RbK ($\sigma_f$), and distinguishable 
scattering between LiK-RbK ($\sigma_d$).
We have used $D=5.29\times10^{-6}$cm ($10^3a_0$); this value of $D$ can easily 
be achieved for each system and is discussed below.
Note the divergence of the cross sections containing s-wave contributions, and
the Wigner suppression of $\sigma_f$.
In fig. \ref{sigma}(b) we show the dimensionless quantity $k\sigma$
as a function of $Dk$. $k\sigma$ is proportional to the scattering rate, and 
$(Dk)^2$ is twice the ratio of the
kinetic energy and the dipolar energy of the system, $2E/(d^2/D^3)$.
The symbols are the data from (a), but the energy range has been extended to 
show the low energy trend of the scattering.
In fig. \ref{sigma}(b) we also show three analytic methods to estimate the 
scattering cross section: Born Approximation (solid blue line),  threshold 
s-wave scattering (black triangles), and semi-classical (solid green line);  
each of these estimates is detailed below.

\begin{figure} 
\vspace{2mm} 
\centerline{\epsfysize=48mm\epsffile{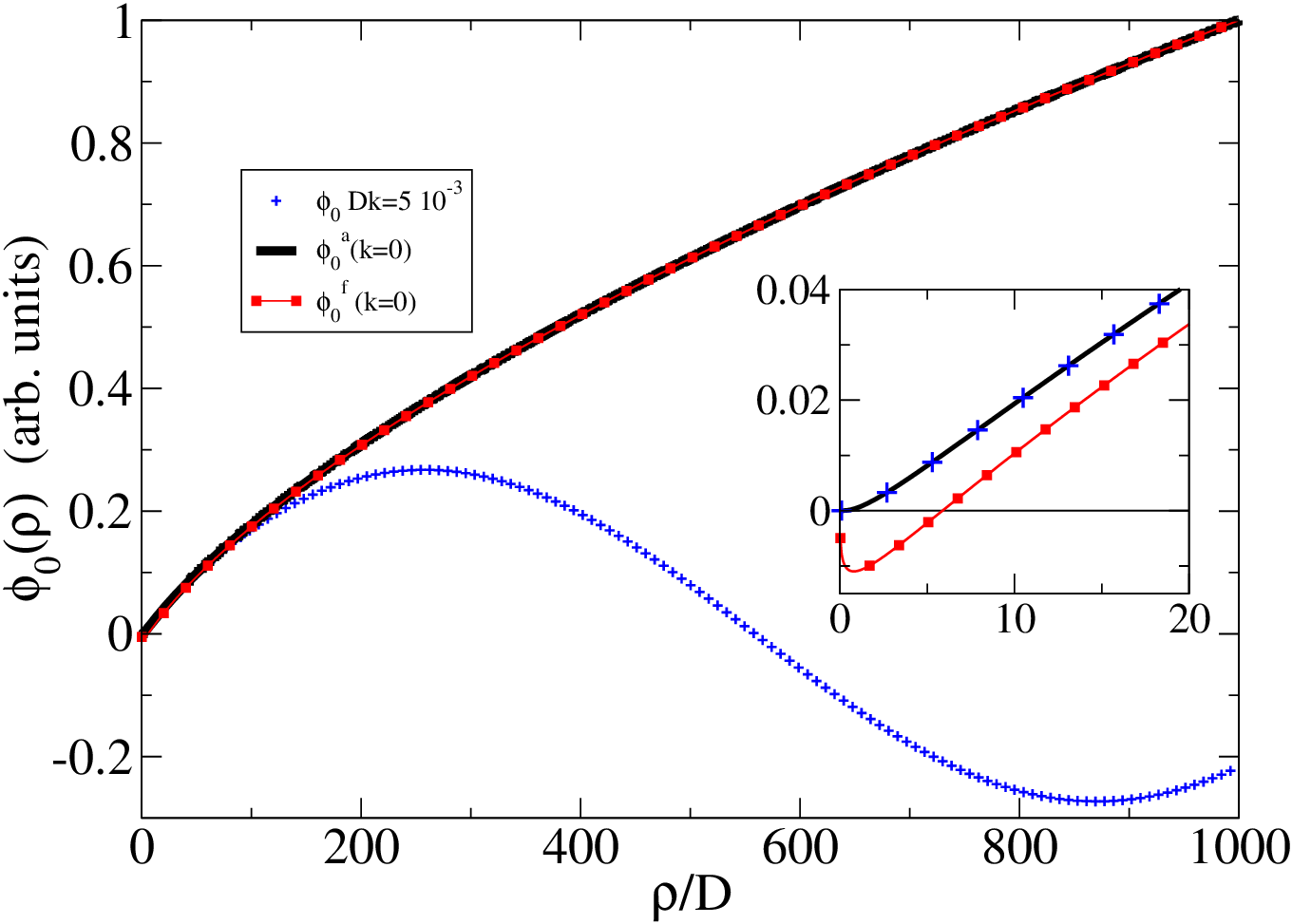}}
\caption{(Color Online) The analytic s-wave zero energy wavefunction 
(solid black line) and the zero energy free solution, (red squares) 
are shown.  Also plotted is a wavefunction in the threshold regime 
($Dk=5\times10^{-3}$, blue x), which agrees well with $\phi_0^a$ at small 
$\rho/D$.  The inset shows these wavefunctions close to the origin.  
}\label{wave}
\end{figure}

To obtain the threshold scattering properties of the 2D dipolar problem, 
we use an analytic zero energy ($k=0$) solution to Eq. (\ref{TISE}):
\begin{eqnarray}
&&\phi_m^a(\tilde\rho)=2\sqrt{\tilde\rho} 
\left[K_{2m}\left(\sqrt{8/\tilde\rho}\right)
+CI_{2m}\left(\sqrt{8/\tilde\rho}\right)\right]\label{solution}
\end{eqnarray}
where $K_{2m}/I_{2m}$ are modified Bessel functions.  
$C$ is determined by the boundary condition: $\phi_0^a(\tilde\rho_0)=0$.  
Considering the s-wave case in the limit of small $\tilde\rho_0$, 
$C$ becomes small and is proportional to $e^{-2\sqrt{8/\tilde\rho_0}}$.  
In this limit 
$\phi_0^a(\tilde\rho)=2\sqrt{\tilde\rho}K_{2m}(\sqrt{8/\tilde\rho})$ and 
this wavefunction is plotted in Fig. \ref{wave} as a solid black line.  We have 
also plotted $\phi_0$ (blue x) for $Dk=5\times10^{-3}$, and it agrees well 
with $\phi_0^a$ at small $\rho/D$, as is shown by the inset. 

\begin{figure} 
\vspace{2mm} 
\centerline{\epsfysize=48mm\epsffile{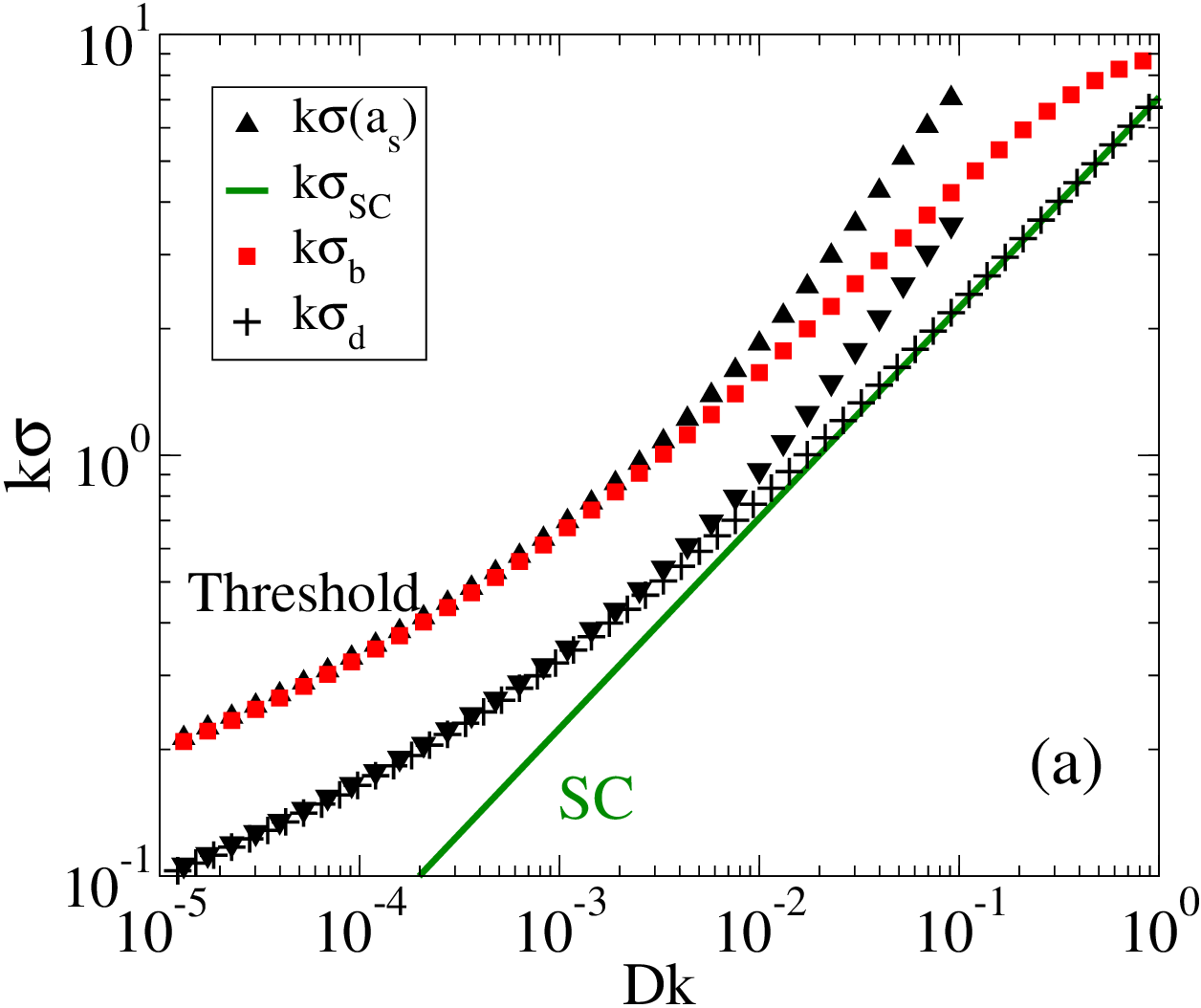}} 
\vspace{4mm} 
\centerline{\epsfysize=46mm\epsffile{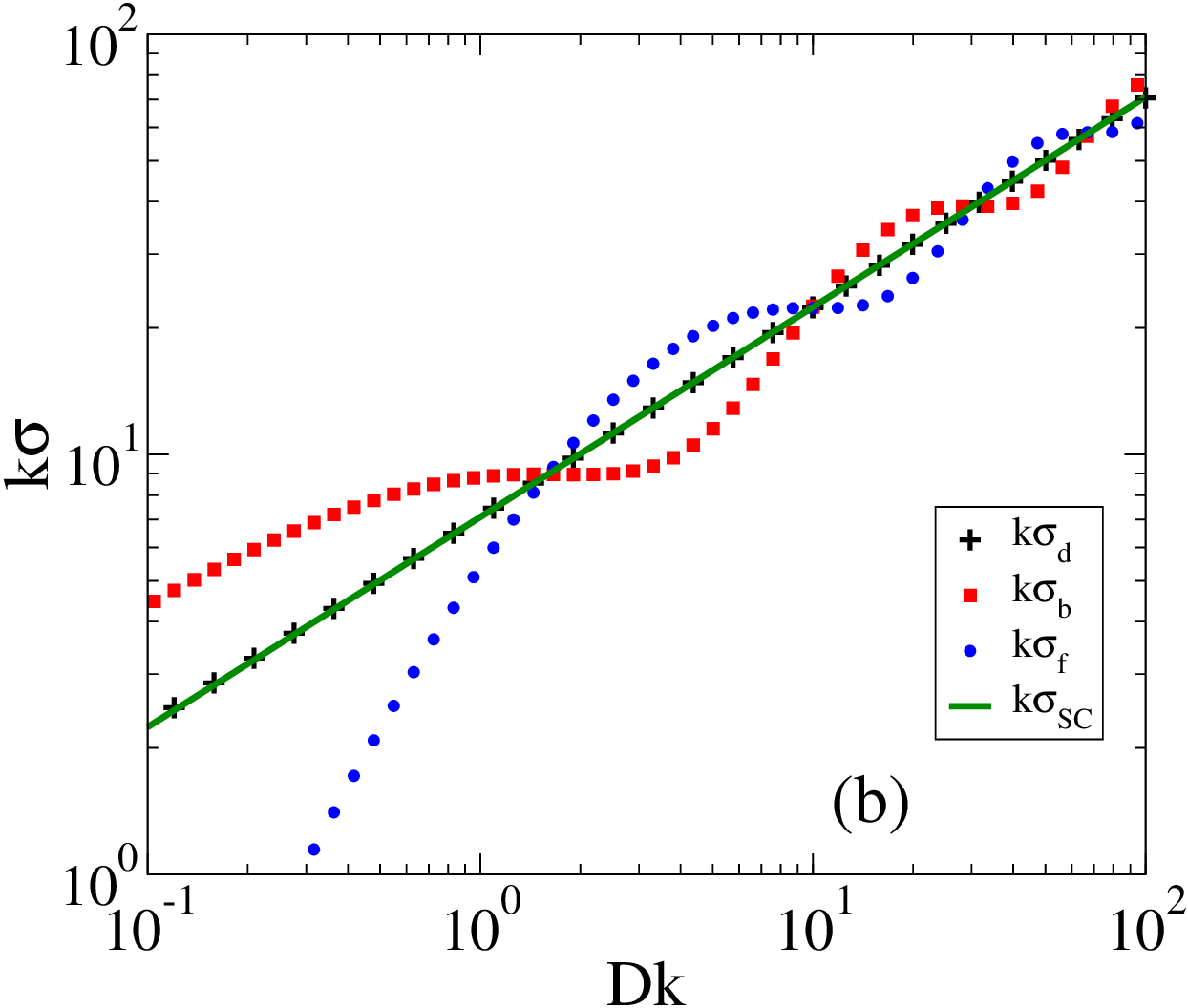}} 
\caption{(Color Online) (a) The low energy s-wave scattering for both 
$k\sigma_d$ (black +) and $k\sigma_b$ (red squares) with
$k\sigma_{SC}$ (solid green line), and the $k\sigma_{hw}(a_s)$ (black triangle)
for both the bosonic and distinguishable cases.
(b) The high energy behavior of $k\sigma_f$ (blue circles), 
$k\sigma_b$ (black +), and $k\sigma_d$ (red squares) with 
$k\sigma_{SC}$ (solid green line).  The $\sqrt{Dk}$ scaling of $k\sigma$ 
is clearly shown.
}\label{swaveEIK}
\end{figure}

To exploit the analytic solution, we find the form of the free 
solution, $\phi_0^f$, which is analogues to the zero energy 3D wavefunction: 
$r\psi_0^{3D}(r)=r-a_s$, which has the property: $\psi_0^{3D}(a_s)=0$.  
In 2D the scattering length is defined as where zero energy free solution is 
zero: $\phi^f_0=\sqrt{\tilde\rho}\ln(\tilde\rho/\tilde a_s)$ \cite{verhaar}.
To extract the s-wave scattering length, we take the large $\tilde\rho$ 
limit of $\phi_0^a$ and find:
\begin{eqnarray}
&&{\phi_0^a}\rightarrow \sqrt{\tilde\rho}
\left[-2\gamma-\ln(2)+\ln(\tilde\rho)\right]=
\sqrt{\tilde\rho}\ln(\tilde\rho/\tilde a_s)
\nonumber\\
&&\tilde a_s={a_s\over D}\equiv e^{2\gamma+\ln(2)}\sim6.344.
\label{asc}
\end{eqnarray}
where $\gamma\sim0.577...$ is Euler's constant.  Note $\phi_0^a$ in the large 
$\tilde\rho$ limit is the zero energy free solution, and the free solution
is plotted in Fig. \ref{wave} as red squares.  The inset shows the wavefunctions
near the origin.  The threshold cross section for 2D dipolar scattering can be
approximated by the cross section of a hard wall at $a_s$, 
i.e. $\phi_0^f(a_s)=0$:
\begin{eqnarray}
\sigma_{th}=\sigma_{hw}(a_s).\label{swave}
\end{eqnarray}
This is a remarkable result in both its simplicity and accuracy.  In Fig. 
\ref{swaveEIK}(a) we have plotted 
$k\sigma_d$ (black +), $k\sigma_b$ (red squares),
$k\sigma_{hw}$ (black triangle) for both the bosons and distinguishable case, 
and $k\sigma_{SC}$ (solid green line) as a function of $Dk$.  
From this figure we see Eq. (\ref{swave}) reproduces the scattering well in 
the threshold regime when $Dk<10^{-2}$.  

To predict the threshold scattering of non-zero partial waves we use
the Born Approximation (BA) \cite{adhikari,cross}, and find 
\begin{equation}
\sigma_{m}^{BA}={4\over k}{(Dk)^2\over(m^2-1/4)^2}.\label{ba}
\end{equation}
This offers a very good estimate of the non-zero partial waves scattering 
cross sections.  For identical fermions, the BA gives 
$k\sigma_f^{BA}\sim29(Dk)^2$.
We have plotted the BA (blue solid line) in Fig. \ref{sigma}(b) for $k\sigma_f$
(blue circles).  In the small $Dk$ limit, the BA reproduces the $k\sigma_f$ 
well.  In Fig. \ref{sigma} (a), we see that the fermionic cross section 
($\sigma_f$) is Wigner suppressed, as expected \cite{Sadeghpour}.  This is in 
contrast to 3D where all non-zero partial wave become energy independent as 
energy goes to zero \cite{NJP}.  
The analytic solution presented by 
Ref. \cite{bo} produces the results in Eq. (\ref{ba}) when applied to 2D.

To estimate the transition out of the threshold regime for $m=1$, we can 
also equate the centrifugal and dipolar terms and find the energy of that length
scale to be roughly $Dk\sim0.3$. This is a reasonable estimate of when the 
scattering leaves the threshold regime, and is seen in 
Fig. \ref{sigma} (b).  The transition from the threshold regime does require
higher energy collisions. $D$ can be larger, which activates more partial waves.

The BA breaks down when $Dk$ becomes large and we enter into 
the semi-classical regime.  In this limit the scattering is no longer dominated
by a single partial wave, and to understand the scattering in the 
semi-classical limit one must obtain the total cross section.  To do this, 
we estimate the scattering amplitude with the 2D Eikonal 
approximation \cite{eikonal}:
$f_{eik}={-ik2\over\sqrt{2\pi}}\int_{-\infty}^\infty db \cos(kb\theta)
[e^{i\delta_{eik}}\sin(\delta_{eik})]$, where the phase is
$\delta_{eik}={-1\over2k}\int_b^\infty \rho d\rho 
{U(\rho)\over\sqrt{\rho^2-b^2}}$, and $b$ is the impact parameter.  For the 
dipolar interaction $U(\rho)=2D/\rho^3$, the phase shift is $-Dk/(bk)^2$.  
Using the optical theorem, $\sigma=\sqrt{8\pi}/k Im[f(0)]$ \cite{adhikari}, 
we find the total semi-classical cross section to be
\begin{eqnarray}
\sigma_{SC}={4\over k}\sqrt{\pi Dk}.\label{sc}
\end{eqnarray}
To illustrate the universal behavior of the scattering,
in Fig. \ref{swaveEIK} (b)
we plot $k\sigma_b$ (red squares), $k\sigma_d$ (black +), 
$k\sigma_f$ (blue circles), and analytic estimate from Eq. (\ref{sc}):
$k\sigma_{SC}$ (green solid line) as a function of $Dk$.   
At large $Dk$, $k\sigma$ clearly shows the semi-classical $\sqrt{Dk}$ scaling. 
$k\sigma_b$ and $k\sigma_f$ oscillate out of phase about $k\sigma_{d}$. 
 Fig. \ref{swaveEIK} (b) shows that the semi-classical approximation offers a 
very good estimate of the scattering when $Dk>1$.

We have characterized the long range scattering as a function of
$Dk$ in the $\tilde\rho_0\ll1$ limit.  This ensured that deviations from 
universal behavior due to the short range interaction were negligible.  
But this leads to an important question: when can the scattering be impacted 
by the short range interaction?  We have surveyed this question and have found 
that for $\tilde\rho_0$ less than 0.1 the short range plays little to no role 
in the scattering, even for $Dk\ll1$.
When $\tilde\rho_0$ exceeds this value, then the short range can resonantly 
impact the scattering and spoil the universal threshold
scattering.  Once $\tilde\rho_0>1$, the scattering cannot be characterized by 
dipolar scattering alone, and knowledge of the short range is required.  A 
more thorough treatment of the short range impact on 2D dipolar scattering 
will be presented in the future.

\begin{figure} 
\vspace{3mm} 
\centerline{\epsfysize=46mm\epsffile{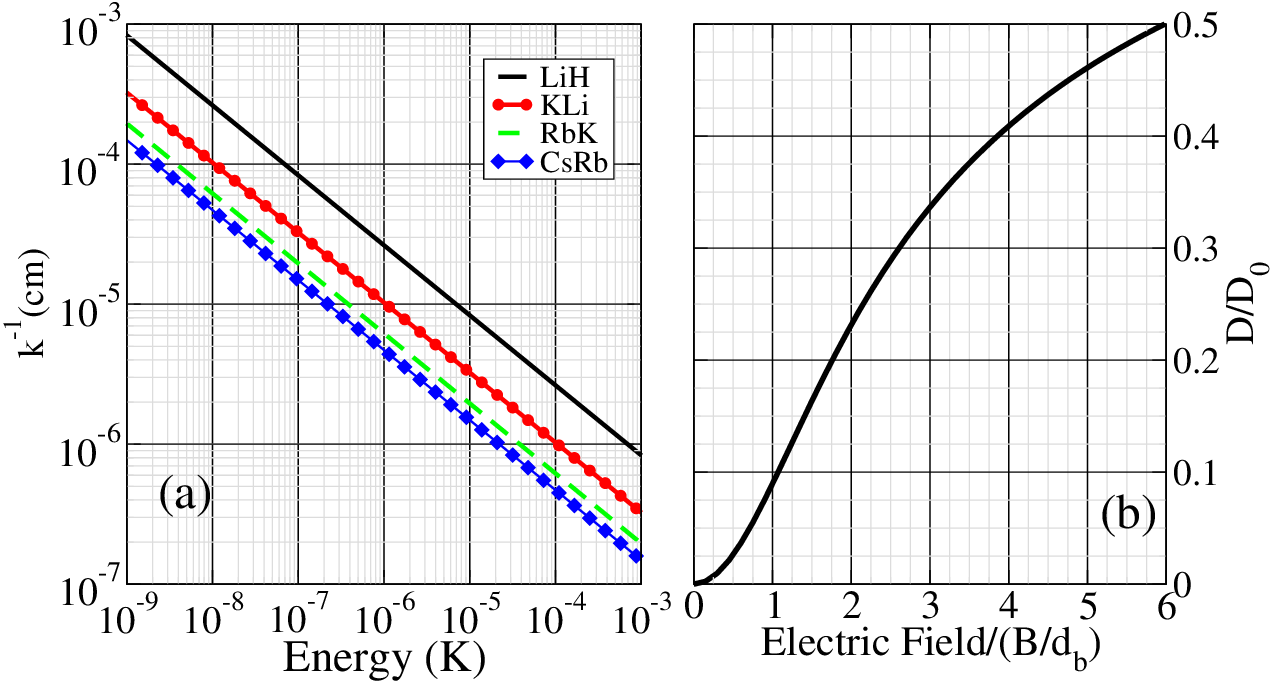}} 
\caption{(Color Online) (a) $k^{-1}$ is shown as a function of energy
for several ${}^1\Sigma$ molecules:
LiH (black solid line), KLi (red circles), RbK (green dashed line)
and RbCs (blue squares).  (b) $D/D_0$ is shown as a function of electric 
field.   With these two curves one is able to 
estimate the scattering cross section for physically reasonable energies and 
electric field values.
}\label{energy}
\end{figure}

An aim of this work is to facilitate estimates of the cross
section for 2D dipolar scattering.  This can be done with 
Fig. \ref{energy}.  In (a) we have plotted  $k^{-1}$ as a function of 
energy for for LiH (black solid line), KLi (red squares), RbK (green dashed 
line), and RbCs (blue circles).  With this and the value of $D$ one can 
readily determine the scattering character and estimate a value of the 
scattering cross section.  

To estimate $D$, we use Fig. \ref{energy} (b) where we have plotted $D/D_0$ 
as a function of electric field. $D_0$ is the length scale obtained by using 
the bare dipole moment, $d_b^2 m/\hbar^2$.  
For the ${}^1\Sigma$ molecules \cite{gspm,carr}, the polarization of the 
molecules depends on the strength of the external electric field (${\cal E}$) 
in relation to the rotation energy of the molecules; therefore the 
field can be characterized by $B/d_b$.   
Fig. \ref{energy} (b) shows $D/D_0$ which depends on the induced dipole 
moment squared ($d^2$) as a function of electric field over $B/d_b$. 
Note that the $D$ is only a fraction of this length.  A few examples are: 

\begin{tabular}{ccccccc}
&RbCs&RbK&KLi&LiH&LiCs\\\hline
$D_0/10^{-5}$ (cm)&28&3&45&20&317\\
${\cal E}_0$ (kV/cm)&0.8&3.9&4.2&74&0.5\\\hline
\end{tabular}

Many others are given in Ref. \cite{roudnev}.    For mixtures, such as LiK-RbK,
we need to determine the induced dipole moment of each molecule at a 
particular field: $D=d_1d_2m_r/\hbar^2$.
To offer a concrete example, consider fermionic RbK at $1\mu K$, where 
$k^{-1}\sim5\times10^{-6}$cm.
In Fig. \ref{sigma} (a) we used $D=5.29\times10^{-6}$, this requires
${\cal E}\sim1.25{\cal E}_0=4.9kV/cm$.  Then $Dk\sim1$, and
we use the semi-classical estimate and find $k\sigma\sim4\sqrt{\pi Dk}\sim7$.
This estimate yields $\sigma\sim3.5\times10^{-5}$cm, which is quiet close to the
numerical value in Fig. \ref{sigma}(a).

In conclusion, we have characterized the long range 2D dipolar scattering and
have offered several means to estimate the scattering cross section.
In the {\it universal threshold regime} we have found an analytic 
solution to the interacting 2D radial Schr\"{o}dinger equation.  
From this solution, we have extracted the scattering length for 
2D dipolar scattering: $a_s=6.344D$.   
Additionally we have explored the {\it semi-classical universal} scattering, 
and the results are that the scattering cross section 
scales as $\sqrt{D/k}$, from Eq. (\ref{sc}). 
This regime can be reached by not only increasing temperature, but also
by increasing $D$.  The strong dipolar interaction 
activates many partial waves to scattering and the scattering can be  
semi-classical even at ultracold temperatures.
An important application of this work will be the use of 
$a_s$ in many body theories to correctly account for the dipolar interaction
\cite{petrov,kan,wang}. 
Future work will explore the scenario where $\hat d\cdot\hat\rho\ne0$ and the   
inclusion of transverse confinement in the scattering.

\begin{acknowledgments}
The author gratefully acknowledges support from the Australian Research Council
and partial support from NSF through ITAMP at Harvard University
and Smithsonian Astrophysical Observatory. The author thanks
H. Hu and X.-J. Li for incredibly enlightening discussions.
\end{acknowledgments}
\bibliographystyle{amsplain}

\end{document}